\documentclass[10pt,a4paper]{article}
\usepackage[utf8]{inputenc}
\usepackage{amsmath}
\usepackage{amsfonts}
\usepackage{amssymb}
\usepackage{graphicx}
\usepackage{cite}
\usepackage{hyperref}
\title{Comment on ``Casimir effect in a weak gravitational field: Schwinger's approach"}
\author{A.P.C.M. Lima$^a$\footnote{augustopcml@gmail.com}, G. Alencar$^a$ and R.R. Landim$^a$}
\date{{\it ${}^a$Departamento de F\'{\i}sica, Universidade Federal do Cear\'{a}-
Caixa Postal 6030, Campus do Pici, 60455-760, Fortaleza, Cear\'{a}, Brazil.}\\
\bigskip
\today}
\begin{document}
\maketitle
\begin{abstract}
We show that the statement in F. Sorge [Class. Quant. Grav. \textbf{36}, no.23, 235006 (2019)] that the Casimir effect receives second order corrections due to gravity is not consistent. We remark especially on the tracing of the proper time Hamiltonian, where the correct procedure is to use the eigenfunctions and eigenvalues of the covariant D'Alembertian. After some cancellations we find that the value of the functional $W[0]$ is the same as obtained by Sorge. However, we argue that the proper vacuum energy density carries extra space-time volume terms that cancel over the gravitational correction, returning  to the same expression as in Minkowski space-time.
\end{abstract}

\section{On the tracing procedure for the proper time Hamiltonian}

As shown by Schwinger in \cite{SchwingerI,SchwingerII}, the Casimir energy \cite{casimir} can be obtained from the functional $W[0]\equiv i \ln(Z[0])$ , which can be written as
\begin{equation}\label{w}
W[0]=\frac{i}{2}\int \frac{ds}{s}Tr[\exp(-i s \hat{H})],
\end{equation}
where $\hat{H}$ is the proper time Hamiltonian. The trace contained in (\ref{w}) can be simply expressed in condensed notation as
\begin{equation}\label{trace}
Tr[\exp(-i s \hat{H})]=\sum_n e^{-i s \lambda_n},
\end{equation}  
where $\lambda$ are the eigenvalues of $\hat{H}$ in any given representation. The proper Hamiltonian can be interpreted as an differential operator acting on the space of scalar fields obeying Dirichlet conditions $\psi(z=0)=\psi(z=L)=0$. Given an complete set of eigenfunctions $\psi_n$ with normalization
\begin{equation}\label{normal}
\int dv_x\psi_n^*\psi_m=\delta_{nm},
\end{equation}
where $dv_x$ is the invariant space-time volume element, the trace (\ref{trace}) can be expanded as \cite{Parker:2009uva}
\begin{equation}\label{trace2}
Tr[\exp(-i s \hat{H})]=\int dv_x\sum_n e^{- i s \lambda_n}|\psi_n(x)|^2.
\end{equation}

For flat space-time, we have $\hat{H}\equiv \square$, and we can use the well known normalized eigenfunctions 
\begin{equation}\label{efflat}
\psi_n=\frac{1}{(2\pi)^{3/2}}\sqrt{\frac{2}{L}}\sin(n\pi z/L)\exp[i(\omega t-k_\perp x_\perp)].
\end{equation}
Thus, we have
\begin{equation}
W[0]=\frac{i}{2}\sum_n\int d^4x d\omega d^2k_\perp\frac{ds}{s}|\psi_n(x)|^2\exp[-is(\omega^2-k_\perp^2-n^2\pi^2/L^2)],
\end{equation}
this is equivalent to the expression presented in \cite{Sorge:2019ldb}. In curved space-time, the same expression (\ref{w}) is valid, we just have to adjust the elements in (\ref{trace2})(e.g. space-time volume element, eigenfunctions and eigenvalues of $\hat{H}$).

With the above arguments in mind, next we would like to make a few remarks in the calculation from \cite{Sorge:2019ldb}. The author splits the total value of $W[0]$ in its flat space-time value plus a gravitational correction coming from the modified proper time Hamiltonian
\begin{equation}
W[0]=W_{flat}[0]+\delta W[0].
\end{equation}
Even though the operator whose trace is to be evaluated in the first term is the same as in Minkowski case, in the second term
\begin{equation}
\delta W=-\frac{i}{2}\int\frac{ds}{s}Tr[4i\gamma s(z+is\partial_z)\nabla^2 e^{-is\hat{H}_0}],
\end{equation}
the field modes (\ref{efflat}) do not represent eigenfunctions of the operator in square brackets, and the subsequently $z$ dependent values to be integrated do not correspond to formal eigenvalues. Metric terms are missing in the space-time integration, although this is canceled out by the use of the flat modes. Notice also that the proper time Hamiltonian used is simplified by a factor of $(1-2\gamma z)$, which needs to be accounted for in the eigenvalue equation. 

In the next section we will perform an explicit calculation of the $W[0]$ functional value given by
\begin{equation}\label{w2}
W[0]=\frac{i}{2}\int d^4x\sqrt{-g}\sum_\lambda\int\frac{ds}{s}e^{-is\lambda}|\psi_\lambda(x)|^2,
\end{equation}
where the $\lambda$ and $\psi_\lambda$ are to obtained from the eigenvalue equation
\begin{equation}\label{eigen}
\tilde{\square}\psi_\lambda(x)=\lambda\psi_\lambda(x),
\end{equation}
and $\tilde{\square}$ is the generalized D'Lambertian.

\section{Casimir energy density}

We begin with the eigenvalue equation (\ref{eigen}), which for the metric
\begin{equation}\label{metrica}
ds^2=(1+2\gamma z)dt^2-(1-2\gamma z)(dx^2+dy^2+dz^2),
\end{equation}
can be written as
\begin{equation}
[(1-2\gamma z)\partial_t^2-(1+2\gamma z)\nabla^2]\psi_\lambda=\lambda\psi_{\lambda}.
\end{equation}
As in \cite{Sorge:2005ed}, we expand the solutions as
\begin{equation}
\psi_{n,k,\omega}(x)=A_{n,k,\omega}\chi_n(z) e^{i(\omega t-k_\perp x_\perp)},
\end{equation}
leading to
\begin{equation}
[(1-2\gamma z)\omega^2+(1+2\gamma z)(k^2-\partial_z^2)]\chi_n(z)=-\lambda_{n,k,\omega}\chi_n(z).
\end{equation}
This equation can be rearranged into
\begin{equation}
\partial_z^2\chi-a z\chi+b\chi=0,
\end{equation}
where $a=(4\omega-2\lambda)\gamma$ and $b=(\omega^2-k_\perp^2+\lambda)$. Solutions to the above equation can be found in analogy to the mode solutions in \cite{Sorge:2005ed} (in \cite{Lima:2019pbo} an explicit expression of these modes as perturbations to the flat case solutions is shown). Also from \cite{Sorge:2005ed}, imposing the boundary conditions on the resulting modes leads to
\begin{equation}
b-\frac{aL}{2}\simeq\frac{n^2\pi^2}{L^2},
\end{equation}
which can be used to obtain the eigenvalues
\begin{equation}\label{lambda}
\lambda=\left(k_\perp^2+\frac{n^2\pi^2}{L^2}\right)(1+\gamma L)-\omega^2(1-\gamma L),
\end{equation}
for $\lambda=0$, we recover the mode frequencies originally presented in \cite{Sorge:2005ed}. Notice that if we had simplified the proper-time Hamiltonian as in \cite{Sorge:2019ldb}, we would have obtained different values.

Then, with the insertion of (\ref{lambda}) and the regularization parameter from \cite{farina},  equation (\ref{w2}) becomes
\begin{align}
W^{(\nu)}[0]=&\frac{i}{2}\sum_n\int d^4xd^2k_\perp d\omega ds s^{\nu-1}\sqrt{-g}|\psi_{n,k,\omega}|^2\\
&\times\exp\left[-is\left(k_\perp^2+\frac{n^2\pi^2}{L^2}\right)(1+\gamma L)+is\omega^2(1-\gamma L)\right].
\end{align}
Proceeding with the calculation as usual (use condition (\ref{normal}) to perform space-time integrations), we find
\begin{equation}\label{w3}
W^{(0)}[0]=(1+\gamma L)^{-1+3/2}(1-\gamma L)^{-1/2}W_{flat}^{(0)}[0]=-(1+\gamma L)\frac{AT\pi^2}{1440L^3}.
\end{equation}
This coincides with the result in \cite{Sorge:2019ldb}. 

It is important to notice that the functional $W$ is directly related to the vacuum to vacuum transition rates as measured in the world line of the static observer and is constructed as an scalar quantity. Remember the similarly invariant mean proper Casimir energy density constructed in \cite{Sorge:2005ed}, defined as
\begin{equation}
\bar{\epsilon}=\frac{1}{V_p}\int d^3x \sqrt{h}u^\mu u^\nu \langle 0\vert T_{\mu\nu}|0\rangle,
\end{equation}
where $h$ is the induced metric determinant and $V_p$ the proper volume of the cavity. In the present approach this quantity can be obtained by factoring out the space-time 4-volume(remember $W[0]$ also contains the time factor), so:
\begin{equation}\label{density}
\bar{\epsilon}=\frac{W[0]}{V_p^{(4)}},
\end{equation}
where $V_p^{(4)}\equiv \int d^4x\sqrt{-g}$. This is in contrast with the original proposal from \cite{Sorge:2019ldb}, where the denominator in the above equation is just $V^{(4)}=ALT$, using coordinate rather than proper parameters. Thus
\begin{equation}\label{energy}
\bar{\epsilon}=-(1+2\gamma L)\frac{\pi^2}{1440L^4}=-\frac{\pi^2}{1440L_p^4},
\end{equation}
where $L_p$ is the proper length $L_p=\int_0^L dz\sqrt{-g_{33}}$. This recovers the result from \cite{Lima:2019pbo}, indicating no correction to the Casimir energy to order of $\gamma$.

\section{Conclusion}

Recently some of the present authors have shown that the treatment of Ref. \cite{Sorge:2005ed} about gravity corrections to Casimir energy was not correct. In fact, a consistent computation leads instead to  the absence of gravitational corrections in Casimir energy up to second order  in $[M/R]^2$ \cite{Lima:2019pbo}. However, Sorge proposes to re-obtain its original result from \cite{Sorge:2005ed}, but now using the Schwinger's approach\cite{Sorge:2019ldb}. In the present work we revisit this calculation and show again that the correction is null. For this we use the same method and considerations used in the original paper \cite{Sorge:2019ldb}. 

The discussion boils down to two main aspects of the original calculation. First one is the tracing procedure for the proper time hamiltonian, where we argue that the treatment of eigenfunctions and eigenvalues is inappropriate. Although the values obtained for the $W[0]$ coincide in this case, there is no guarantee that the same will hold for different space-time metrics.

The second and most important point, as it is the one that effectively changes the final result, is the extraction of the Casimir energy density from $W[0]$. The sought after quantity as originally defined in \cite{Sorge:2005ed} is an invariant mean proper energy density, but the definition used in \cite{Sorge:2019ldb} is not invariant. Therefore, it has instead to be related to $W[0]$ by (\ref{density}), considering the proper space-time volume element. This recovers the result from \cite{Lima:2019pbo}, i.e. null second order gravitational corrections, providing further proof that the results from \cite{Sorge:2005ed, Sorge:2019ldb} are inconsistent. Thus, the analysis provided here is of critical importance in correcting the previous papers.

With the present result the authors seek to solve the current debate, as it is of relevance in the context of vacuum weight experiments \cite{Avino:2019fdq}, where the space-time to be considered is of the form (\ref{metrica})(i.e. Schwarzschild weak field approximation), and thus directly related to this discussion. The present authors also generalized the above results to other metrics. The result is out of the scope of this manuscript  and can be found in a separate paper, where we show that the gravitational corrections are also null for a larger class of space-times \cite{Lima:2020igm}. Therefore the present paper reinforces the rather unexpected fact that, at least to second order, gravity does not change the Casimir energy.

\section*{Acknowledgements}

The authors would like to thank Alexandra Elbakyan and sci-hub, for removing all barriers in the way of science.

We acknowledge the financial support  by Conselho Nacional de Desenvolvimento Cient\'ifico  e Tecnol\'ogico(CNPq) and Funda\c{c}\~ao Cearense de Apoio ao Desenvolvimento Cient\'ifico e Tecnol\'ogico(FUNCAP) through PRONEM PNE0112-00085.01.00/16.

\end{document}